\documentclass[aps,prb,twocolumn,superscriptaddress]{revtex4-1}
\usepackage{graphicx} % For importing images
\usepackage{float} % extra commands for placing figures
\usepackage{comment} % allows to comment out text

\newcommand{\Na}[1][0.8]{Na$_{#1}$CoO$_2$}		%Na0.8CoO2		\Na{}
\newcommand{\NaCa}[2]{Na$_{#1}$Ca$_{#2}$CoO$_2$}	%NaxCayCoO2		\NaCa{x}{y}
\newcommand{\NaCaa}{Na$_{0.65}$Ca$_{0.1}$CoO$_2$}	%Na0.65Ca0.1CoO2\NaCaa
\newcommand{\NaCab}{Na$_{0.57}$Ca$_{0.14}$CoO$_2$}	%Na0.57Ca0.14CoO2\NaCab
\newcommand{\NaCac}{Na$_{0.55}$Ca$_{0.2}$CoO$_2$}	%Na0.65Ca0.1CoO2\NaCac
\newcommand{\NaCas}{Na$_{0.7}$Ca$_{0.1}$CoO$_2$}	%Na0.7Ca0.1CoO2	\NaCas
\newcommand{\ve}[1]{\mathbf{#1}}    %Vector symbol	\ve
\newcommand{\vd}[1]{\mathbf{#1'}}	%Vector' symbol	\vd
\newcommand{\SG}{$P6_3/mmc$}					%P63mmc			\SG
\begin{document}

\title{Divacancy superstructures in thermoelectric calcium-doped sodium cobaltate}

% Authors
\author{D.G. Porter}
\email[]{Dan.Porter@diamond.ac.uk}
\affiliation{Department of Physics, Royal Holloway, University of London, Egham TW20 0EX, UK}
\author{M. Roger}
\affiliation{Service de Physique de l'Etat Condensé, (CNRS/MIPPU/URA 2464), DSM/DRECAM/SPEC, CEA Saclay, P.C. 135, F-91191 Gif Sur Yvette, France}
\author{M.J. Gutmann}
\affiliation{ISIS, Science and Technology Facilities Council, Rutherford Appleton Laboratory, Didcot OX11 0QX, UK}
\author{S. Uthayakumar}
\affiliation{Department of Physics, Royal Holloway, University of London, Egham TW20 0EX, UK}
\author{D. Prabhakaran}
\author{A.T. Boothroyd}
\affiliation{Department of Physics, University of Oxford, Clarendon Laboratory, Parks Road, Oxford, OX1 3PU, United Kingdom}
\author{M.S. Pandiyan}
\affiliation{Department of Physics, Royal Holloway, University of London, Egham TW20 0EX, UK}
\author{J.P. Goff}
\affiliation{Department of Physics, Royal Holloway, University of London, Egham TW20 0EX, UK}

\date{\today}

\begin{abstract}
We have grown single crystals of \NaCa{x}{y} and determined their superstructures as a function of composition using neutron and x-ray diffraction. Inclusion of Ca$^{2+}$ stabilises a single superstructure across a wide range of temperatures and concentrations. The superstructure in the Na$^+$ layers is based on arrays of divacancy clusters with Ca$^{2+}$ ions occupying the central site, and it has an ideal concentration \NaCa{4/7}{1/7}. Previous measurements of the thermoelectric properties on this system are discussed in light of this superstructure. \NaCa{4/7}{1/7} corresponds to the maximum in thermoelectric performance of this system.
\end{abstract}

%\keywords{}

\maketitle

\section{Introduction}
Sodium cobaltate (\Na[x]) has emerged as a material of exceptional scientific and technological interest due to its potential for thermoelectric applications that convert waste heat into electricity or allow solid state refrigeration\cite{RefWorks:4,Wang2003,RefWorks:60,RefWorks:7}. In \Na[x], the removal of sodium ions creates holes in the cobalt layers. The Seebeck coefficient and power factor both increase with $x$, as the concentration of holes decreases\cite{RefWorks:7}. Above a concentration of $x\approx0.85$, however, phase coexistence occurs with the insulating $x=1$ phase, causing a reduction in the power factor. The substitution of a divalent ion for sodium decreases by one the number of holes in the cobalt layer. This offers the possibility to lower the hole concentration and enhance further the thermoelectric performance. There are already promising results on polycrystalline samples doped with Ca and Sr, with an increase in the power factor over the comparable pure compound\cite{RefWorks:32,RefWorks:33,RefWorks:8}.

Neutron diffraction studies have demonstrated the central role played by superstructure on the thermoelectric properties of \Na[x] \cite{RefWorks:10,RefWorks:16}. The structure of \Na[x] comprises layers of conducting cobalt-oxygen octahedra separated by layers of sodium ions. The sodium ions can sit on one of two inequivalent sites; Na $2b$ sites directly between successive cobalt ions, or displaced from these position on Na $2d$ sites. The $2d$ sites are preferentially occupied due to a small energy cost for sitting directly above cobalt arising from short-range repulsion. Vacancies in the sodium layer have been found to form multi-vacancy clusters at certain fractional concentrations. These clusters form by promoting sodium ions onto the $2b$ sites, creating rattling cages that can reduce thermal conduction through the material\cite{Extra:02}. Coulomb interactions cause the vacancy clusters to arrange themselves into long range patterns, causing rings of superlattice peaks in diffraction images. 

The electrostatic potential generated by these superlattices templates the Coulomb landscape in the cobalt-oxygen layer, generating confined electronic pathways in the layers\cite{RefWorks:10}. The combination of electron confinement and rattling cage-ions creates an ideal environment for the thermoelectric effect.

Here we present the results of an x-ray and neutron diffraction study of superstructures in Ca-doped \Na[x]. We show that the Ca stabilises a single superstructure over a wide range of temperature and composition, and we use the results to gain an understanding of how Ca-doping affects the thermoelectric properties.

\section{Methods}
Single crystal boules of various compositions of \NaCa{x}{y} were grown using the floating zone technique \cite{RefWorks:36,RefWorks:135}. Small, high quality crystallites were cleaved from the boules and studied using x-ray diffraction with a molybdenum source, Xcalibur diffractometer (Agilent instruments) at Royal Holloway. A Cryojet5 (Oxford Instruments) was used for temperature control and precise temperatures were calibrated by replacing the sample with a thermistor of similar size. A large, single grain of \NaCas{} was cleaved from its boule and studied using neutron diffraction on SXD \cite{RefWorks:48} at ISIS. 

The hexagonal reflections from the diffraction experiments were integrated using standard instrument methods, and properties of the average structure were refined using the program Jana2006\cite{RefWorks:117}. A reverse Monte Carlo (RMC) program was written to determine the vacancy ordering within the sodium layer and to establish the position of calcium ions. RMC uses simulated annealing  to minimise the chi-squared difference between the calculated intensities arising from a given set of atomic positions in the unit cell, and the experimental data. The positions of Ca and Na ions are changed with respect to the Co lattice, using the Metropolis algorithm in the Canonical ensemble. Small displacements of all atoms, with respect to their original positions, are also allowed. When the fictitious temperature is slowly decreased, the configuration corresponding to the absolute minimum of this chi-square difference can be reached. Further details on this method can be found in Refs. [\cite{RefWorks:76,Extra:01}]. 

\section{Diffraction Experiments}
\begin{figure}[ht]
	\centering
	\includegraphics[width=0.5\textwidth]{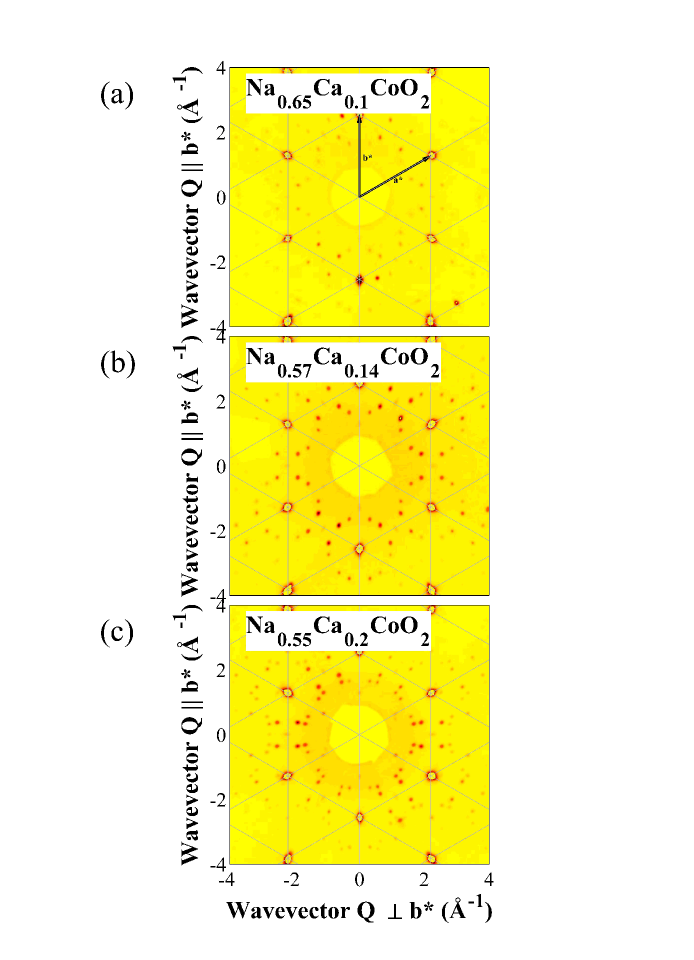}
	\caption{(Colour online) Reciprocal space cuts through the $(h,k,0)$ layer for different compositions of \NaCa{x}{y}, measured by single-crystal x-ray diffraction. \textbf{(a)} \NaCaa{} shows a 12-fold ring of satellite peaks around the principal Bragg reflections which can be indexed on a hexagonal grid with spacing $a^*/7$. \textbf{(b)} \NaCab{} displays the same superlattice structure, but with stronger intensity. \textbf{(c)} \NaCac{} also displays the $a^*/7$ superlattice peaks, plus additional peaks that lay on a $a^*/4$ lattice.} \label{fig:1}
\end{figure}

Three compositions of the calcium doped system were grown using the floating zone technique: \NaCaa{}, \NaCab{} and \NaCac{}. The samples were grown as large 4--6 cm boules which were cleaved to find small, high quality crystallites. These were screened with the x-ray diffractometer and high quality samples were measured with longer exposures and greater coverage at several temperatures. Figure \ref{fig:1} shows reciprocal space cuts in the $(h,k,0)$ plane from a single-crystal sample of each composition. All three samples exhibit rings of 12 satellite reflections around the principal hexagonal Bragg peaks. These additional peaks lay on the vertices of a hexagonal grid with spacing $a^*/7$ and can be indexed using a supercell with lattice vectors: 
\[\vd{a}=2\ve{a}-\ve{b}\] \[\vd{b}=\ve{a}+3\ve{b}\]  \[ \vd{c}=\ve{c} \]
These supercell vectors index one of the two possible superlattice domains, where the second domain can be generated by a reflection in either the $\{1,0,0\}$ or $\{1,1,0\}$. The intensities of the superlattice reflections indicate equal populations of these domains. No broadening of the peaks was detected within instrumental resolution and, therefore, the superstructures were found to be ordered long range. The positions of these superlattice peaks are not consistent with any of the diffraction patterns observed previously in pure \Na[x], signifying that a new superstructure has been found. 
All three samples have additional reflections halfway between the largest Bragg reflections due to contamination by $\lambda/2$, in addition to this the \NaCac{} sample displays further reflections between the primary satellites. These additional reflections lay at the vertices of an $a^*/4$ hexagonal grid and can be indexed by a supercell with lattice vectors:
\[\vd{a}=2\ve{a}\] \[\vd{b}=\ve{a}+2\ve{b}\]  \[ \vd{c}=\ve{c} \]
These propagation vectors were first observed for \Na[0.5] \cite{RefWorks:25}, implying a coexistence between the \Na[0.5] superstructure and the new $a^*/7$ superstructure.

% Figure 1 XRD hk0 3 compositions

\begin{figure*}[htp]
	\centering
	\includegraphics[width=0.45\textwidth]{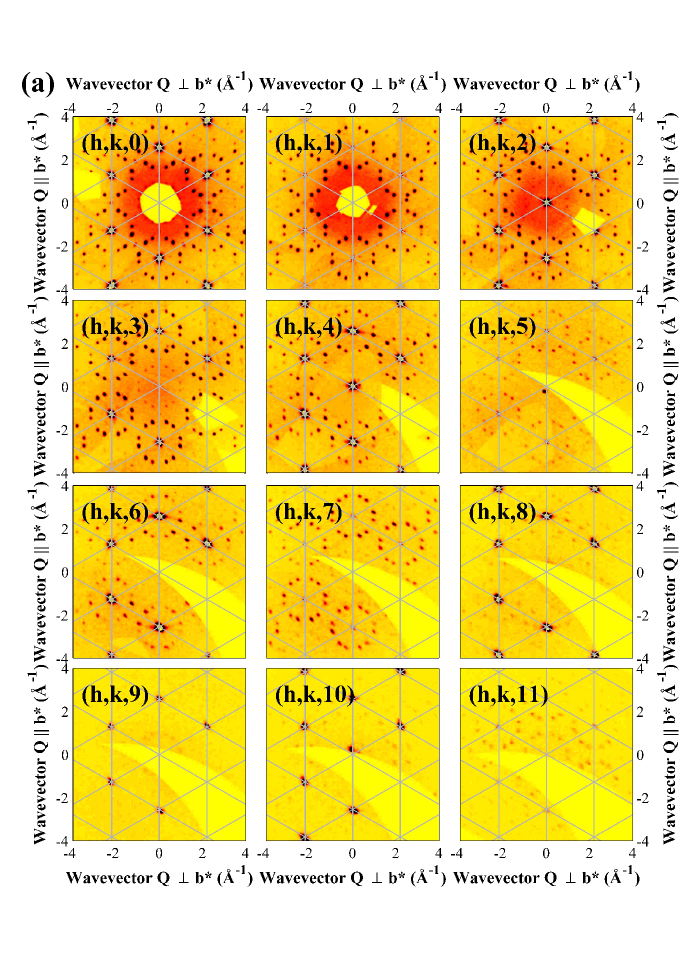} \includegraphics[width=0.45\textwidth]{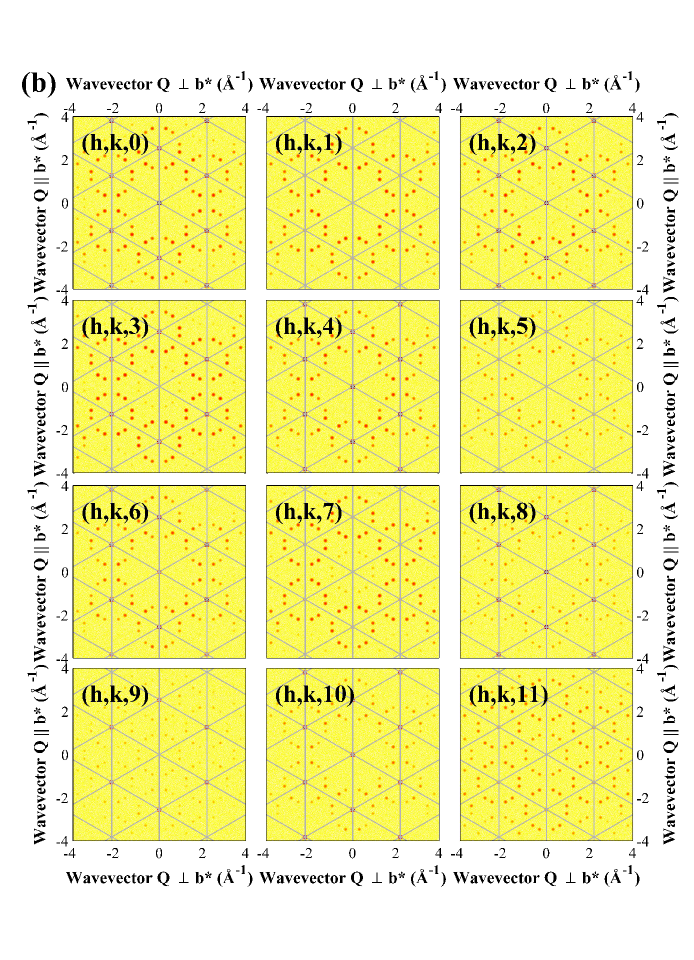}
	\caption{(Colour online) \textbf{(a)} Cuts through the reciprocal space volume for \NaCab{} at integer $l$ values, illustrating the variation of intensity in the superlattice peaks with $l$. Gaps in the data are a result of missing coverage. \textbf{(b)} Calculated intensities from the model generated by reverse Monte-Carlo, for comparison with \textbf{(a)}.} \label{fig:2}
\end{figure*}

% L-variation
The intensities of the satellite peaks vary significantly for different $l$ planes as shown in Fig. \ref{fig:2}. Satellites occur primarily in 12-fold rings around the $\{1,0,l\}$ reflections, with the modulation of peak intensity within the rings varying between even and odd planes. At high $l$, the peak intensity is diminished due to the x-ray form factor, however peaks can still be observed and show that satellites around the $\{1,1,l\}$ reflections gain intensity, especially in the $(h,k,11)$ plane. The increased satellite intensity and reduced intensity in the principal reflections in the $(h,k,11)$ plane is particularly striking and it is related to the fact that the $c$-lattice parameter is approximately 11 times the distance between the cobalt and oxygen planes.

% Figure 2 LDependence & Model

\begin{figure}[ht]
	\centering
	\includegraphics[width=0.5\textwidth]{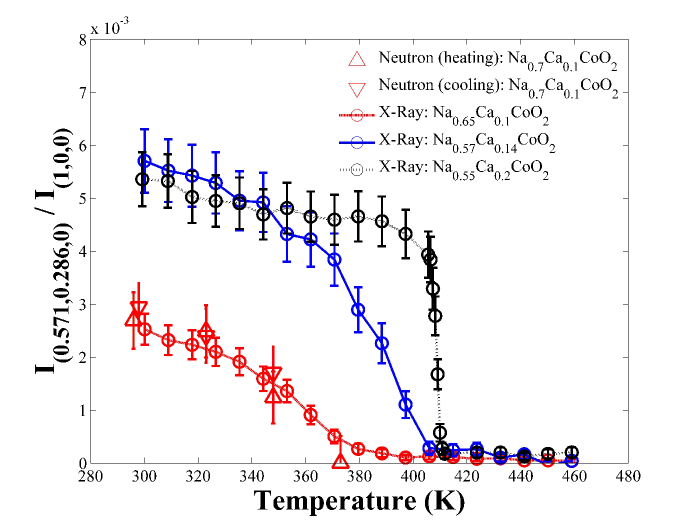}
	\caption{(Colour online) Temperature dependence of the superlattice peaks for three compositions of \NaCa{x}{y}. Each peak intensity is normalised by a close principal hexagonal reflection, therefore indicating the relative intensity variation between compositions. Fitting an order parameter to each dataset gives disorder temperatures of $T_d=379\pm6K$ for \NaCaa{}, $T_d=401\pm2K$ for \NaCab{} and $T_d=410\pm10K$ for \NaCac{}. The neutron data points indicate measurements during heating and cooling, showing no sign of hysteresis behaviour and are consistent with the x-ray measurement.} \label{fig:3}
\end{figure}

% Temperature dependence
% Updated ``We did not observe any variation...'' for Referee 1, comment 2, June 2014
The diffraction patterns were measured as a function of temperature between 90 and 460 K. It was found that the satellite peaks do not change below room temperature, though they do eventually disappear at high temperature, as illustrated in Fig. \ref{fig:3}. The disappearance of the satellite peaks is attributed to the sodium layers becoming disordered. We did not observe any variation of the peak widths in this temperature range within our experimental resolution. The ordering temperature increases as the calcium concentration increases, and the order parameter indicates a sharper transition for the $y=0.2$ sample. The $a^*/4$ peaks did not vary at all within this temperature range.

% Figure 3 Temp Dependence
% New version for Referee 1, comment 3, June 2014

The peak intensities in Fig. \ref{fig:3} are normalised against the intensity of a nearby principal hexagonal reflection in each case. The \NaCaa{} intensity is much weaker than the other compositions and it has a lower disorder temperature, with a fitted order parameter giving $T_d=379\pm6 K$. There is little difference between the initial intensity or the disorder temperature of \NaCab{} and \NaCac{}, with $T_d=401\pm2 K$ and $T_d=410\pm10 K$ respectively. This suggests that the $a^*/7$ satellites in both samples originate from a phase with a common calcium composition. The additional calcium in the \NaCac{} sample is, therefore, likely to be part of the coexisting phase.

% SXD Neutron measurements
A boule of \NaCas{} was screened on SXD. A large single grain with intense, sharp principal Bragg peaks was cleaved from the boule and longer exposures of several hours were taken at 350K and 40K. It was not possible to grow large enough single grains for neutron diffraction studies of the superstructures with higher calcium concentrations.

\begin{figure}[ht]
	\centering
	\includegraphics[width=0.5\textwidth]{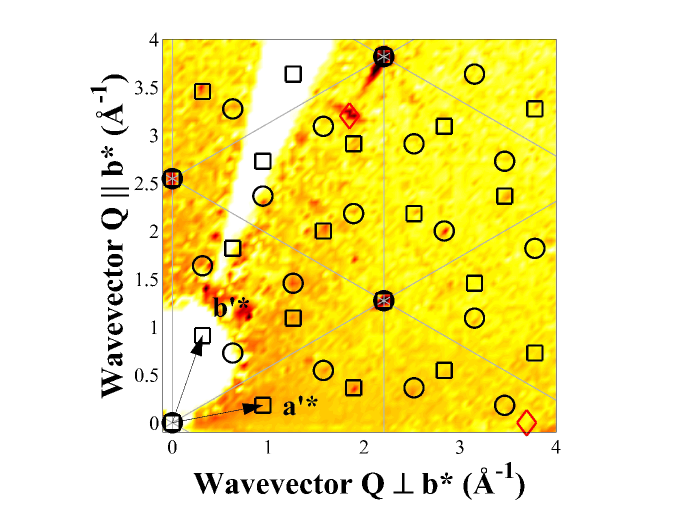}
	\caption{(Colour online) A cut in the $(h,k,0)$ plane for \NaCas{} SXD data at 40K. Satellite peaks appear in the same positions as found in the x-ray data, indicating that this is a bulk phase. Black circles and squares distinguish the two domains for the superlattice, and the red diamonds mark the \{220\} reflections of the epitaxial CaO impurity. Other non-indexed peaks are a result detector anomalies and noise resulting at low detector angles.} \label{fig:6}
\end{figure}

Superlattice peaks can be distinguished in the data by their intensities having hexagonal symmetry. Satellite peaks were observed with the same propagation vectors used to index the $a^*/7$ superlattice observed with x-rays, indicating that this phase forms in the bulk material, as shown in Fig. \ref{fig:6}. Several shorter exposures were performed at increasing temperatures and the disorder transition observed was consistent with the \NaCaa{} measurement performed with x-rays, see Fig. \ref{fig:3}. The superlattice pattern did not change as a function of temperature, indicating a single phase below the ordering transition temperature.

% Figure 4 SXD hk0

% CaO impurity
% Updated ``Comparing relative intensities...'' for Referee 1, comment 4, June 2014
% Edited ``By comparing the integrated intensities...'' for Referee 1, comment 2, July 2014
In addition to the satellites, further peaks were observed close to the $\{110\}$ principal hexagonal reflections. The d-spacing of these peaks was $\approx3.70$ \AA, which matches the $\{220\}$ reflection of calcium oxide (CaO). By comparing the integrated intensities for several reflections with calculated structure factors, and averaging over domains, the relative proportion of CaO in \NaCas{} was determined to be 0.8\% with a statistical error of 0.2\%. For this system to produce a hexagonal pattern of points as found in these results, the cubic $[111]$ direction of the CaO crystal would have to be parallel to the hexagonal $[001]$ of \NaCas{}, and the cubic $[110]$ direction of CaO would be parallel to the hexagonal $[100]$ of \NaCas{}. In this direction the CaO structure can be described as successive hexagonal layers of calcium and oxygen ions with an ABC stacking sequence. The hexagonal environment of the main phase allows the CaO impurity to grow epitaxially within the \NaCas{} boule.

% Refinement
\section{Single Crystal Refinements}
The principal hexagonal reflections in the multi-composition x-ray datasets were integrated using the three-dimensional (3D) profile method. The results of the refinements, as recorded in Table \ref{tab:refNaCa}, show that the sodium occupancies are consistently lower than the nominal values. The calcium occupancies are much closer to the nominal values however, which is consistent with loss of the more volatile sodium ions during growth. For \NaCab{} the fitted Ca occupancy agrees with the nominal composition for the proposed \NaCa{4/7}{1/7} superstructure. This is consistent with the lack of any observed coexisting phases in this sample.

% Table 1: Refinements
\begin{table*}[ht]
\small
\centering
\begin{tabular}{lccccccccccc} \hline
					&			&			& \multicolumn{2}{c}{Occupancy} & \multicolumn{4}{c}{$U_{iso}$}	&\\
Target Composition	& a	(\AA)	& c (\AA)	& Na 		& Ca		& Na		& Ca		& Co  & O & $R_w$	\\ \hline
\NaCaa 				& 2.8398(6)	& 10.746(2)	& 0.56(1) 	& 0.12(1) 	& 0.019(2)  & 0.013(6) 	& 0.0066(5)	& 0.006(1)	&7.84\%	\\
\NaCab				& 2.8403(4)	& 10.756(1)	& 0.529(5)	& 0.138(5)	& 0.010(1) 	& 0.005(2)	& 0.0049(3)	& 0.0058(6)	&3.96\% \\
\NaCac 				& 2.8362(4)	& 10.761(2)	& 0.516(7)	& 0.172(9)	& 0.013(1)  & 0.014(3) 	& 0.0062(4)	& 0.0050(7)	&7.39\% \\ \hline
\end{tabular}
\caption{Refinement of \NaCa{x}{y} principal Bragg reflections from x-ray diffraction data at 300K. The same model was used in each case: Space group: \SG, Positions: Co $(0,0,0)$, O $(\frac{1}{3},\frac{2}{3},0.0930(8))$, Na $(\frac{2}{3},\frac{1}{3},0.25)$, Ca $(0,0,0.25)$. Co and O occupancies were fixed at full occupancy.}
\label{tab:refNaCa}
\end{table*}

% Updated ``A full description'' for Referee 1, comment 2, June 2014
The thermal parameters were refined as isotropic spheres. The data in Table \ref{tab:refNaCa} are consistent with increased static disorder away from the optimum composition. The ADPs for the sodium ions on the $2d$ sites are comparable to those obtained for the pure system\cite{Extra:02}, whereas those of the divalent calcium ions on the $2b$ sites are significantly lower. A full discussion of ADPs in terms of the lattice vibrations is beyond the scope of the present paper.

The R-factors for these refinements show that there is excellent agreement between the refined models and the experimental data. Other refinements were attempted that changed the assumption that Ca ions occupy $2b$ sites and Na ions sit on $2d$ sites. These resulted in unrealistic concentrations of calcium or negative thermal parameters, and these models were, therefore, rejected.

% RMC
\section{Reverse Monte Carlo Simulations}
Intensities from the superlattice reflections from \NaCab{} were obtained from the room temperature x-ray data by indexing a single superlattice domain and excluding the principal hexagonal reflections. The superlattice reflections arise purely from the superstructure, whereas the principal reflections are only slightly modified, and they are also affected by other factors such as multiple scattering, or coexisting domains and phases. The propagation vectors used to index the rings of superlattice reflections converts to a real space supercell of seven hexagonal sodium cobaltate unit cells, and the concentrations from the refinements meant that four sodium ions and one calcium were added to each layer. RMC calculations were performed using this set of intensities and supercell. Each iteration of the RMC program involves discrete hopping movements of the sodium and calcium ions to vacancy sites within the plane, as well as continuous movements of any ion by small increments in any direction. Simulated annealing of the system helped to avoid a false minimum solution, initially allowing ions to hop freely then slowly decreasing the fictitious temperature and reducing the probability of allowing movements that would produce a higher $\chi^2$. 

\begin{figure}[h]
	\centering
	\includegraphics[width=0.4\textwidth]{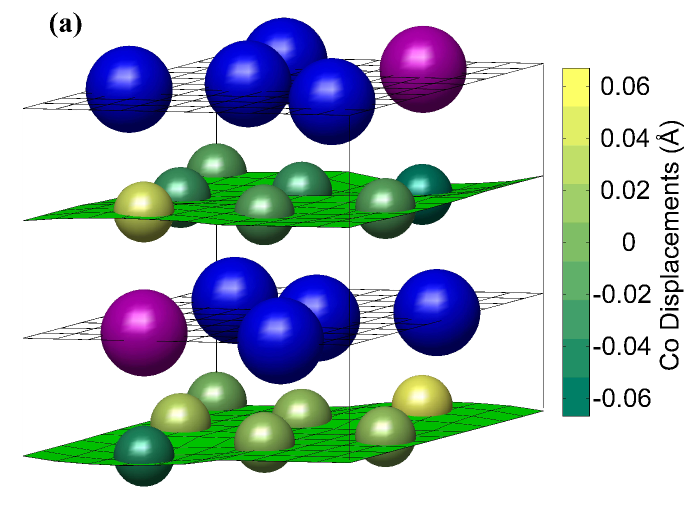} \\ \includegraphics[width=0.4\textwidth]{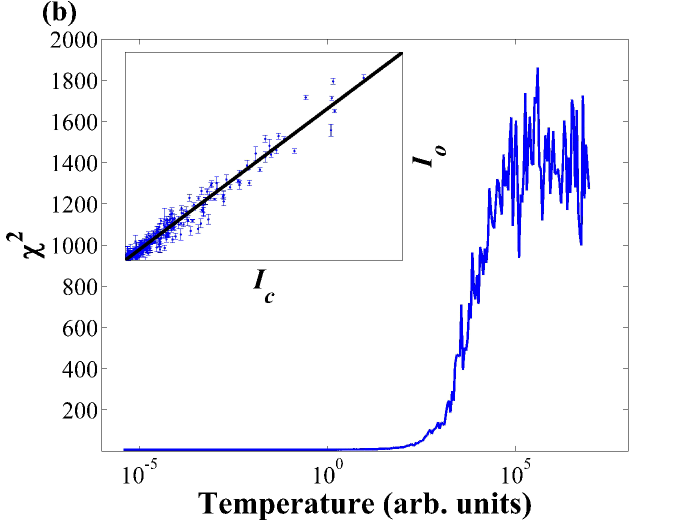}
	\caption{(Colour online) Results from the RMC calculation of \NaCab{} x-ray data. \textbf{(a)} The resultant structure, where blue balls are sodium ions, purple are calcium ions, the other balls are cobalt ions, where dark green is a large negative distortion and light green is a large positive distortion. Oxygen ions have been omitted for clarity. \textbf{(b)} Annealing of $\chi^2$ during the calculation. Inset: Final comparison of calculated ($I_c$) and experimental ($I_o$) intensities, where the black line displays $I_c = I_o$.} \label{fig:4}
\end{figure}

% Solution
The RMC calculations led to the robust solution illustrated in Fig. \ref{fig:4}, with the calcium ion on a $2b$ position surrounded by sodium ions sitting on $2d$ sites. In the second sodium/calcium layer the same pattern emerges, however, the calcium ion is translated away from the location of calcium on the first layer. This translation relates to a significant reduction in $\chi^2$. The fit for this model had a $\chi^2$ per degree of freedom of 5.82 and an $R_w$ value of 12.9\%. Figure \ref{fig:2}(b) shows the calculated x-ray patterns for this solution indicating good agreement with the experimental data in Fig. \ref{fig:2}(a) across all $l$ planes, including the variation of the satellite peak intensity between even and odd planes. The resulting structure has the space group $P2_1/m$ with unique axis $c$, atomic positions are given in Table \ref{tab:rmcNaCa}.

% Figure 5 Reverse Monte Carlo
% Updated caption for Referee 1, typo 1, June 2014

% Table 2: Refined supercell positions
\begin{table}[ht]
\small
\centering
\begin{tabular}{lcccc} \hline
Name & $x$ & $y$ & $z$ & Site\\ \hline
  Co1 & 0 & 0 & 0 & 2a \\
  Co2 & 0.1453(2) & 0.7156(2) & 0.00453(18) & 4f \\
  Co3 & 0.2850(2) & 0.4295(2) & -0.00029(17) & 4f \\
  Co4 & 0.4266(2) & 0.1421(2) & 0.00147(17) & 4f \\ 
  O1 & 0.0462(11) & 0.2425(12) & 0.0926(9) & 4f \\
  O2 & 0.2436(11) & 0.1951(11) & 0.5877(9) & 4f \\ 
  O3 & 0.1948(11) & 0.9557(12) & 0.0977(9) & 4f \\
  O4 & 0.3759(11) & 0.9031(11) & 0.5922(9) & 4f \\
  O5 & 0.3284(11) & 0.6601(12) & 0.0955(8) & 4f \\
  O6 & 0.5271(11) & 0.6205(11) & 0.6018(9) & 4f \\
  O7 & 0.8966(11) & 0.5176(11) & 0.0853(9) & 4f \\
  Na1 & 0.2334(14) & 0.1921(15) & 0.25 & 2e \\
  Na2 & 0.0592(14) & 0.2410(15) & 0.75 & 2e \\
  Na3 & 0.3799(14) & 0.8966(15) & 0.25 & 2e \\
  Na4 & 0.5226(14) & 0.6177(15) & 0.25 & 2e \\
  Ca1 & 0.1486(8) & 0.7151(8) & 0.75 & 2e \\
\end{tabular}
\caption{Atomic coordinates of the \NaCab{} structure, determined by RMC. The structure has the monoclinic space group $P2_1/m$ with unique axis $c$. The lattice parameters are $a'=7.5177(6)$\AA, $b'=7.5201(6)$\AA, $c=10.7564(9)$\AA, $\gamma=119.986(8)^\circ$.}
\label{tab:rmcNaCa}
\end{table}

% Distortions
The RMC solution also exhibits ordering in the distortions of the cobalt plane, with cobalt ions moving away from the closest calcium (which is sitting directly above the cobalt). This form of distortion was predicted using a Coulomb model for the pure system\cite{RefWorks:10}. The magnitude of the distortion is smaller for the system doped with calcium since the decreased hole concentration reduces the electrostatic potential for the multi-vacancy clusters. Figure \ref{fig:4}(a) depicts the calcium doped divacancy structure with cobalt distortions emphasized for clarity. It was not possible to resolve the variation in the Co-O bond lengths using our x-ray data and, therefore, we attempted to use the neutron data.

% Neutron RMC
RMC simulations using the neutron diffraction data were less sensitive to the superlattice, mainly due to a lower signal-to-background. The same divacancy superstructure was obtained, but the smaller scattering contrast between the calcium and sodium ions made it difficult to locate the positions of the calcium ions.

\begin{figure}[h]
	\centering
	\includegraphics[width=0.5\textwidth]{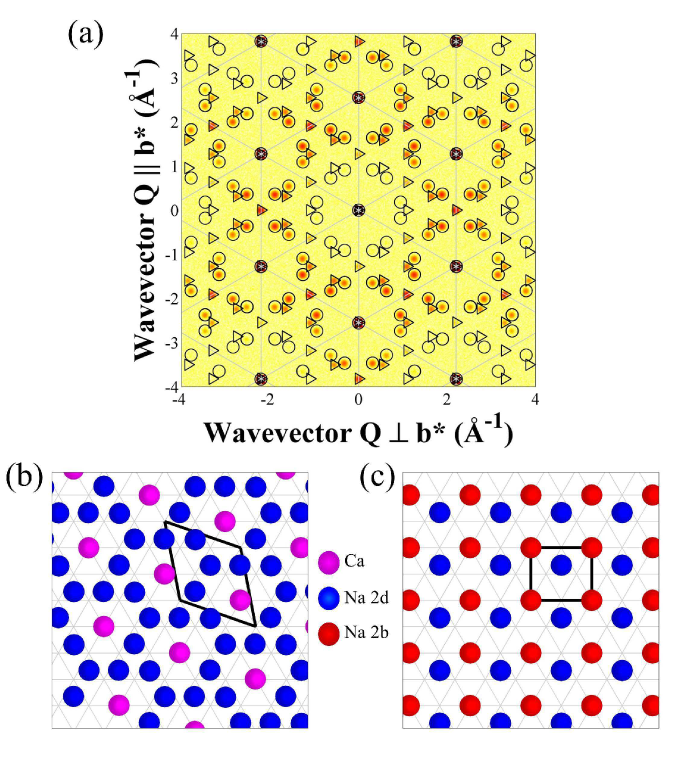}
	\caption{(Colour online) \NaCac{} can be described by a coexistence of the \NaCab{} supercell in \textbf{(b)} and the $x=0.5$ supercell in \textbf{(c)}. The calculation of the $(h,k,0)$ layer in \textbf{(a)} gives good agreement with the x-ray data from Fig. \ref{fig:1}(c). Superlattice peaks generated by the \NaCab{} supercell are indicated with circles, whereas the peaks produced by the $x=0.5$ supercell are marked by triangles.} \label{fig:5}
\end{figure}

% 50% coexistence
For \NaCac{} this divacancy phase coexists with the phase observed for \Na[0.5] as shown in Fig. \ref{fig:5}. In fact, the \Na[0.5] superstructure can also be viewed as stripes of divacancy clusters\cite{RefWorks:10}. In this case it was not possible to perform RMC simulations due to the presence of contamination by higher order diffraction peaks.

% Figure 6 Phases
% updated image for Referee 1, typo 2, June 2014

% Discussion
\section{Discussion}
% Structure - comparison + stability
Analysis of the superlattice observed in all samples of \NaCa{x}{y} indicates the presence of the superstructure with an ideal concentration of \NaCa{4/7}{1/7}. For pure sodium cobaltate the corresponding superstructure has a sodium concentration $x=5/7$. According to calculations of the ground state energies of multi-vacancy structures using a long-range Coulombic model with short-range repulsion, this is the only stable divacancy cluster in this composition range\cite{RefWorks:10}. It is likely that the substitution of the divalent calcium at the centre of the divacancy cluster has a stabilising effect since it lowers the charge of the divacancy cluster. The fact that this superstructure does not change with temperature and forms over a range of sample compositions suggests that it is a particularly stable superstructure.

% Thermoelectric power factor
% Updated ``In these reports...'' for Referee 2, comment 2, June 2014
Previous studies involving the doping of sodium cobaltate with calcium have concentrated on the variation of transport properties relevant to the thermoelectric performance of the material and had not considered the role of the superstructure. This has led to a number of different doping regimes being attempted though none have used exactly the concentration of the superstructure determined here. Kawata \textit{et al.} found that both the electrical resistivity, $\rho$, and the Seebeck coefficient, $S$, increase with increasing calcium concentration\cite{RefWorks:32}. Ono \textit{et al.} found that the increase in both $\rho$ and $S$ was linear, so that overall the power factor, $P=S^2/\rho$, also increased with calcium concentration\cite{RefWorks:33}. However, the power factor reaches a maximum value for \NaCa{0.70}{0.0175} and above this calcium concentration, $P$ decreases. In another composition series, Li \textit{et al.} found that the power factor first increased, and then decreased for \NaCa{0.64}{0.16} and higher calcium concentrations\cite{RefWorks:8}. In these reports the power factor is peaked very close to the optimum concentration for the ideal \NaCa{4/7}{1/7} superstructure. The Seebeck coefficient increases up to this composition. Above this composition, where we find coexistence with an additional insulating phase, the electrical conductivity decreases.

% Updated for Referee 1, comment 3, July 2014
For a full understanding of the thermoelectric performance of this material, the effect of the divacancy superstructure on the thermal transport must also be understood, as this is a key component for calculating the thermoelectric figure-of-merit, $ZT = TS^2/\rho\kappa$, where $\kappa$ is the thermal conductivity. This contribution is beyond the scope of the current work but will be addressed in a future publication.

% Conclusion
\section{Conclusions}
In summary, both x-ray and neutron  diffraction of calcium-doped sodium cobaltate have revealed a previously unobserved superlattice that can be explained by a superstructure with seven unit cells. An RMC program was used to find the ordering in both sodium layers as well as distortions in the cobalt plane in a model-independent manner, reliably generating a divacancy structure with calcium ions at the centre on a $2b$ site, with an ideal concentration \NaCa{4/7}{1/7}. The same superstructure is observed across a broad temperature range and in all compositions studied, though moving away from the optimal concentration for this structure results in weaker superlattice intensities and the emergence of additional phases. Comparison with thermoelectric measurements in the literature indicates that this ideal concentration coincides with the maximum in the thermoelectric power factor for the system.

\begin{acknowledgments}
We thank D. Voneshen for his help. We are grateful for the financial support and hospitality of ISIS. This work was supported by EPSRC grants EP/J011150/1 and EP/J012912/1.
\end{acknowledgments}

\bibliography{thesisBIBLIO}

\end{document}